\input harvmac
\input epsf
\noblackbox

\def\br{\hfill\break}
\def\frac#1#2{\relax{\textstyle {#1 \over #2}}\displaystyle}
\def\cicy#1(#2|#3)#4{\left(\matrix{#2}\right|\!\!
                     \left|\matrix{#3}\right)^{{#4}}_{#1}}
\def\IQ{\relax\,\hbox{$\inbar\kern-.3em{\rm Q}$}}
\def\IB{\relax{\rm I\kern-.18em B}}
\def\IC{{\bf C}}
\def\IP{\relax{\rm I\kern-.18em P}}
\def\IR{\relax{\rm I\kern-.18em R}}
\def\ZZ{\relax\ifmmode\mathchoice
{\hbox{Z\kern-.4em Z}}{\hbox{Z\kern-.4em Z}}
{\lower.9pt\hbox{Z\kern-.4em Z}}
{\lower1.2pt\hbox{Z\kern-.4em Z}}\else{Z\kern-.4em Z}\fi}
\def\Xh{\hat{X}}

\def\h{{1\over2}}
\def\ph{\hat{p}}
\def\d{\partial}
\lref\swI{N. Seiberg and E. Witten, 
{\it Nucl. Phys.} {\bf B426} (1994) 19, hep-th/9407087.}
\lref\swII{
A. Klemm, W. Lerche, S. Theisen, and S. Yankielowicz,
       {\it Phys. Lett.} {\bf B344} (1995) 169, hep-th/9411048;\br
P. Argyres and A. Faraggi, {Phys. Rev. Lett.}
{\bf 74} (1995) 3931, hep-th/9411057.}
\lref\coni{P. Candelas, A. Dale, C. L\"utken and R. Schimmrigk,
{\it Nucl. Phys.} {\it B298} (1988) 493,
P. Greene and T. H\"ubsch, {\it Phys. Rev. Lett.} {\bf 61} 1163,
{\it Commun. Math. Phys.} {\bf 119} (1988) 431,
P. Candelas, P.S. Green and T. H\"ubsch, {\it Phys. Rev. Lett.} {\bf 62}
(1989) 1956; {\it Nucl. Phys.} {\bf B330} (1990) 49.}
\lref\strI{A. Strominger, {\it Nucl.Phys.} {\bf B451} (1995) 96,
hep-th/9504090.}
\lref\gms{B. Greene, D. Morrison and A. Strominger, 
{\it Nucl.Phys.} {\bf B451} (1995) 109, hep-th/9504145} 
\lref\kv{S. Kachru and C. Vafa, {\it Nucl.Phys.} {\bf B450} (1995) 69,
hep-th/9505105.}
\lref\fhsv{S. Ferrara, J. Harvey, A. Strominger and C. Vafa, 
{\it Phys.Lett.} {\bf B361} (1995) 59, hep-th/9505162.}
\lref\klmII{S. Kachru, A. Klemm, W. Lerche, P. Mayr and C. Vafa, hep-th/9508155.}
\lref\klmI{A. Klemm, W. Lerche and P. Mayr, Phys.\ Lett.\ {\bf B357} 
(1995) 313, hep-th/9506112.}
\lref\vawi{C. Vafa and E. Witten, hep-th/9507050}
\lref\aslu{ P. Aspinwall and J. Louis, hep-th/9510234}
\lref\check{V. Kaplunovsky, J. Louis, and S. Theisen, 
{\it Phys.Lett.} {\bf B357} (1995) 71, hep-th/9506110;
I. Antoniadis, E. Gava, K. Narain, and T. Taylor, 
{\it Nucl.Phys.} {\bf B455} (1995) 109, hep-th/9507115.}
\lref\resol{see e.g. W. Fulton, {\it Introduction to Toric
Varieties}, Princeton University Press, Princeton (1993)}
\lref\Candelas{P. Candelas, X. De la Ossa, A. Font, S. Katz, and D.
Morrison, {\it Nucl. Phys.} {\bf B416} (1994) 481, hep-th/9308083.}
\lref\Yau{S. Hosono, A. Klemm, S. Theisen, and S.T. Yau, {\it Comm.
Math. Phys.} {\bf 167} (1995) 301.}
\lref\agm{P. Aspinwall, B. Greene and D. Morrison, 
{\it Nucl.Phys.} {\bf B420} (1994) 184.}
\lref\bsv{M. Bershadsky, V. Sadov and C. Vafa, hep-th/9510225.}
\lref\aspI{P. Aspinwall, hep-th/9510142.}
\lref\vafaFoC{C. Vafa, {\it Nucl. Phys.} {\bf B477} (1995) 262}  
\lref\bcov{M. Bershadsky, S. Ceccotti, H. Ooguri and C. Vafa 
(Appendix of S. Katz), {\it Nucl. Phys.} {\bf 405} (1993) 279,
{\it Commun. Math. Phys.} {\bf 165} (1994) 311}
\lref\hktyII{S. Hosono, A. Klemm, S. Theisen and S.T. Yau
{\it Nucl. Phys.} {\bf B433} (1995) 501}  
\lref\bkk{P. Berglund, S. Katz and A. Klemm, 
    {\it Nucl. Phys. } {\bf B456} (1995) 153, hep-th 9506091}
\lref\bkkII{Work in progress}  
\lref\bbs{K. Becker, M. Becker and A. Strominger,
{\it Nucl. Phys.} {\bf B456} (1995) 130, hep-th/9507158} 
\lref\dais{D.I. Dais, Doctorial Thesis, Bonn 1994, Preprint MPI/94-62.}
\lref\rolf{B. Hunt and R. Schimmrigk, hep-th/9512138;\br
M. Lynker and R. Schimmrigk, hep-th/9511058.}
\lref\hosI{S. Hosono, B.H. Lian and S.-T. Yau, alg-geom/9511001}
\lref\fern{G. Aldazabal, A. Font, L.E. Ib\'a\~nez and F. Quevedo, hep-th/9510093}
\Title{\vbox{
\hbox{CERN-TH-96-02}
\hbox{\tt hep-th/9601014}
}}
{Strong Coupling Singularities and Non-Abelian Gauge}
\vskip-1cm
\centerline{{\titlefont Symmetries in $N=2$ String Theory}}
\bigskip\bigskip
\centerline{Albrecht Klemm and Peter Mayr}
\bigskip\centerline{\it Theory Division, CERN, 1211 Geneva 23,
Switzerland}
\vskip .3in
We study a class of extremal transitions between topological distinct
Calabi--Yau manifolds which have an interpretation in terms of the special
massless states of a type II string compactification. In those cases 
where a dual heterotic description exists the exceptional massless
states are due to genuine strong (string-) coupling effects.
A new feature is the appearance of enhanced non-abelian gauge symmetries 
in the exact nonperturbative theory.
\Date{\vbox{\hbox{CERN-TH-96-02}\hbox{\it {January 1996}}}}

\newsec{Introduction}
The understanding of the nonperturbative dynamics of $N=2$ Yang--Mills
theories in four dimensions \refs{\swI,\swII}\ has subsequently led to a 
substantial improvement of the understanding of nonperturbative
effects in $N=2$ string theories. A striking example is the physical
realization of transitions between topological different Calabi--Yau
manifolds through black hole condensation in the corresponding
type II string compactification \refs{\strI,\gms}. A second highlight
is the conjecture of a $N=2$ type II -- heterotic string duality \refs{\kv,
\fhsv},
which allows to extract the exact nonperturbative vector moduli
space of the heterotic theory from tree-level data of a related type II 
string \klmII. 
A crucial insight is the interpretation of certain singularities
in the exact effective theory as the effect of generically massive
solitonic BPS-states, which become massless at certain points in the 
moduli space. While the effective theory with generically massive states
integrated out diverges when approaching these points, a physical sensible
theory can be obtained by including the additional massless states as
new degrees of freedom.

The conifold singularities \coni~are well understood, both on the
Calabi--Yau side as well as on the heterotic side, however
they represent only one of many types of possible singularities of 
Calabi--Yau spaces \aspI.
A first example of these transitions through 
more general singularities arises very natural as the ``mirror transition'' to 
the standard example of a conifold transition from the quintic 
in $\IP^4$ $\cicy{-200}(\IP^4|5){1,101}$ to the complete 
intersection Calabi--Yau manifold 
$\cicy{-168}(\IP^4 \cr \IP^1|4&1\cr 1&1){2,86}$. 
This conifold transition was interpreted in \gms~in the type IIB 
picture\foot{Our conventions are that we consider a type IIA theory compactified
on the Calabi--Yau manifold $X$ and a corresponding type IIB theory on the mirror
manifold $\Xh$.}, 
where the vector moduli space correspond to the complex structure
deformations (and hence the $(2,1)$--forms) and decreases from $101$ 
to $86$, by the Higgs-breaking mechanism through massless solitonic 
BPS hyper multiplets, which come from three-branes wrapping around 
the vanishing three-cycles. 
The inverse process in the type IIA picture, 
where the vector multiplets correspond to K\"ahler structure deformations 
(and hence (1,1)-forms) and their number decreases from two to one, involves a new 
type of transition, which we will study in detail. It has some similarities
with a generic type of singularities, which appear in the duals of heterotic 
theories and are not continuously connected to the weak coupling regime
\foot{Transitions which occur in the perturbative regime of the heterotic
string connect exclusively $K_3$-fibrations \fern\ and will be studied
in \bkkII}. 
It is the aim of this paper to understand the physics of these two types
of singularities, which have in common the fact that the singular locus
in the moduli space is described by a simple discriminant factor
involving only a certain subset of complex structure moduli $z_s^i$, e.g. for 
$i=1$:
\eqn\disI{\Delta_s= 1- 4\, z_s,}
where $z_s$ is a standard toric coordinate in the complex structure 
moduli space.

As an example consider the two moduli hypersurface 
Calabi--Yau $\cicy-256(\IP_{11226}|12){2,128}$
discussed first in \refs{\Candelas,\Yau}. Its mirror $\Xh$ is described by
the degree twelve polynomial
\eqn\xtwelve{
\hat p=x_1^{12}+x_2^{12}+x_3^6+x_4^6+x_5^2-12\psi x_1x_2x_3x_4x_5-2\phi x_1^6x_2^6
\ .}
The discriminant locus, where the three-point functions diverge,
is described by
\eqn\disI{
\Delta = \Delta_{c} \times \Delta_s = ((1-z_1)^2-z_1^2z_s)\times(1-4z_s) \ ,
}%
where 
\eqn\csmod{
z_1=-{1\over 864}{\phi \over \psi^6},\qquad z_s={1\over4\phi^2} \ .
}%
There is substantial evidence by now \refs{\kv,\klmI,\check,\klmII}\ 
for the conjecture that the type IIA string
theory compactified on $X$ is the dual description of a heterotic string
compactified on $K3 \times T^2$.
According to the identification of \kv, the $T$ and $S$ fields
of the heterotic side should be identified (in the large
$S$/weak coupling regime) with the special coordinates corresponding
to $z_1$ and $z_s$, respectively.  In particular, for large $S$ one has:
\eqn\hetv{ 
z_1={1728\over j(t_1)}+\dots\ ={1728\over j(T)}+\dots\ ,
\ \ \ \ z_s=q_sf(q_1)+\dots = {\exp}(-S)+\dots\ ,}
where $t_s,t_1$ are the special coordinates in the large complex
structure limit and $q_k=\exp(2\pi i t_k)$.
The discriminant factor $\Delta_{c}$ describes a conifold singularity 
tangent to the weak coupling divisor $z_s=0$ in the point $z_1=1 
\Leftrightarrow T=i$, where the heterotic theory has 
a perturbative gauge symmetry enhancement $U(1)\to SU(2)$. In the
exact quantum corrected theory this $SU(2)$ is again broken to $U(1)$,
as in the global supersymmetric theory. In fact it was shown in \klmII,
how the exact nonperturbative string theory reduces to the 
Seiberg--Witten result in the point particle limit. The 
second factor $\Delta_s$ is not connected to the weakly coupled 
regime and a genuine strong coupling singularity. Moreover its
local structure is quite different from the conifold case and of
the type we will consider in the following.

\newsec{Gravitational index}

From the physical point of view we want to interpret the singularity
in the vector moduli space as the effect of additional massless particles.
In \vafaFoC\ it was shown how the net number of additional massless 
multiplets, irrespectively of their gauge quantum numbers, is 
determined from the asymptotic behaviour of the topological 
amplitude $F_1$ defined in \bcov. If $V(z)$ denotes the period 
which vanishes at a singular codimension one locus in the moduli space, 
a leading behaviour 
\eqn\leading{F_1=-{b\over 6} {\rm log} V \bar V}
corresponds to a net contribution of $6 b=n_V-n_H$ additional massless
supermultiplets in the effective field theory (in the following we 
denote $b$ as the gravitational index). Remarkably in all 
examples of conifold singularities considered so far 
\Candelas\hktyII~the value of $b$ equals one, supporting the 
picture developed in \strI. 

Let us explain how the topological limit of $F_1$ can be calculated 
for the type II string on a given Calabi-Yau manifold with 
$h$ deformation parameters. 
A general expression was given in canonical coordinates $t_i= B+i R_i^2$ 
near the large K\"ahler structure (radii) limit 
($R_i\rightarrow \infty$, $z_i\rightarrow 0$) in \bcov
\eqn\FI{F_1^{top}={\rm log}\left[\left(1\over w_0(z_i(t_k))\right)^{3+h-{\chi\over 12}}
{\rm det}\left({\partial z_1\ldots\partial z_h}\over 
{\partial t_1\ldots t_h}\right) f(z_i(t_k))\right],}
where $w_0$ is the locally ($z_i=0)$ unique holomorphic period and the
special coordinates $t_i$ are defined by the mirror map as ratios of 
the geometrical periods as functions of the $z_i$; 
specifically the single logarithmic periods and $w_0$, i.e. 
$t_i={{ w_0 {\rm log}(z_i)+\sigma_i}\over w_0}$, where $\sigma_i$ is a series
in the $z_j$.
The ansatz for the meromorphic function $f(z)=\prod_{i=0}^m \Delta_i^{r_i}
\prod_{j=1}^h z_i^{s_i}$ of the moduli has now to be fixed to yield 
the asymptotic behavior of $F^{top}_1$ at codimension one 
loci were the Picard-Fuchs system has singularities.  
The exponents $s_i$  can be determined from the large radius limits 
$R_i\rightarrow \infty$ $\forall i=1\ldots h$ of $F_1^{top}$  
and the $r_i$ from its interpretation in terms of worldsheet 
instanton contributions \bcov:
\eqn\oneloopexp{{F_1}^{top}=-\!{2 \pi i \over 12} \sum_{i=1}^{h}
t_i\!\int\!c_2 \wedge J_i-\sum_{n_l} \left[2 n^{(1)}_{d_1,\ldots,d_h}\log(\eta(
\prod_{i=1}^{h} q_i^{d_i}))+ {1\over 6} n^r_{d_1\ldots d_h}
\log(1-\prod_{i=1}^h {q_i}^{d_i})\right]+{\rm c.}}
As the invariants $n^r_{d_1,\ldots,d_h}$ for the rational instantons
are known from the expansion of the Yukawa couplings, 
the vanishing (or the knowledge) of elliptic instantons 
contributions $n^{(1)}_{d_1,\ldots,d_h}$ at a given multidegree 
$d_1,\ldots, d_h$ gives further linear equations for the $r_i$.

In the following we give examples\foot{More examples can be 
found in \refs{\Yau,\hktyII,\bkk,\hosI}.} of extremal transitions and calculate
the gravitational index. In the simplest examples a two K\"ahler 
moduli Calabi-Yau space flows to a one K\"ahler 
moduli Calabi-Yau space. The transition occurs, when the K\"ahler 
modulus of the complexified volume $t_s$ shrinks to zero, as will be 
shown in the next section; this 
happens in a codimension one locus in the complexified K\"ahler moduli
space.
As $q_s\rightarrow 1$, independently of $q_1$ the Gromov-Witten invariants 
$n_k^{(g)}$ for the holomorphic embedding of curves of all 
genera $g$ and degree $k$ in the one moduli Calabi-Yau are given by $n^{(g)}_k= 
\sum_s n^{(g)}_{k,s}$, where the $n^{(g)}_{k,s}$ count the number of curves
in the two moduli Calabi--Yau manifold. 
The sum over $s$ turns out to be finite.
There are $n^{(0)}_{0,1}$ isolated rational curves which get 
contracted. According to the interpretation of \bbs\ there will 
be BPS saturated  solitonic states, which arise from two-brane solutions 
wrapping around the contracting curves and become massless for zero
volumn.

The physical picture of the transition as a Higgs effect, described below,
suggest, that the transition requires a tuning of the complex structure 
parameters to a special value too. There is no factorization 
of the two kinds of moduli spaces, because charged fields 
become massless. 
In the case of $K_3$--fibrations \refs{\klmI\vawi\aslu\rolf}\ 
with the identification of the dilaton 
as suggested in \kv, the transition correspond in the
heterotic string picture to a genuine strong coupling effect, which 
occurs as the base $\IP^1$ is shrinking to zero.

$$
\vbox{\offinterlineskip\tabskip=0pt 
\halign{\strut
\vrule#
&\hfil ~$#$~
&\vrule#
&\hfil ~\hfil$#$\hfil~
&\vrule#
&\hfil $\, #$~
&\hfil $#$~
&\hfil $#$~
&\vrule# 
&\hfil $\, #$~
&\hfil $#$~
&\hfil $#$~
&\hfil $#$~
&\hfil $#$~
&\hfil $#$~
&\vrule#
&\hfil ~$#$~
&\vrule#\cr
\noalign{\hrule}
& N_0&&{\rm Extremal\,\, Transition}  &&j:& r_{j}&b_{j}&&i:&s_{i}& 
K_{11i} &c^2_{i}&n^{(0)}_{i,0} &n^{(0)}_{0,i} && n^{(1)}_{i,j} &\cr
\noalign{\hrule}
&1 &&\cicy-168(\IP^4_{1,1,2,2,2}|8){2,86}       &&c:&
-{1\over 6}&-{1\over 6}        &&1:& 
-{17\over 6}&8   &56  &    640& 4 &&n^{(1)}_{i,j}=0,&\cr
&&& \downarrow                                &&s:&
-{5\over 6} &-{2\cdot 2\over 6} &&2:&
         -3&4   &24  &  10032&   &&i<3          &\cr
&&&\cicy-176(\IP^5|4,2){1,89}                 && &
                        &                   &&1:& 
&8   & 56 &       &   &&             & \cr
\noalign{\hrule}
&2 &&\cicy-168(\IP^4\cr \IP^1|4 \,\, 1\cr 1\,\,1){2,86} &&c:&
-{1\over 6}&-{1\over 6}        &&1:& 
-{31\over 6}&5   &50  &    640& 16 &&n^{(1)}_{i,j}=0,&\cr
&&& \downarrow                                &&s:&
             &-{16\over 6} &&2:&
         -3&4   &24  &  10032&   &&i<3          &\cr
&&&\cicy-200(\IP^4|5){1,101}                 && &
                        &                  &&1:& 
&5   & 50 &       &   &&             & \cr
\noalign{\hrule}
&3 &&\cicy-168(\IP^4\cr \IP^1|4\cr 2){2,86} &&c:&
-{1\over 6}&-{1\over 6}        &&1:& 
-{28\over 6}&2   &44  &    640& 64 &&n^{(1)}_{i,j}=0,&\cr
&&& \downarrow                                &&s:&
             &-{2\cdot 32\over 6} &&2:&
         -3&4   &24  &  10032&   &&i<3          &\cr
&&&\cicy-296(\IP^4_{1^4,4}|8){1,149}               && &
                        &                  &&1:& 
&2   & 44 &       &   &&             & \cr
\noalign{\hrule}}
\hrule}$$
\vskip-10pt
\noindent{\bf Table 1:} {\it Extremal transitions from $K_3$-fibrations
Calabi-Yau spaces via the strong coupling singularity. After the 
indication of the pair of models we list the exponents $r_i$ of $F_1^{top}$
\FI~at the conifold $r_c$ and the strong coupling singularity $r_s$ as 
well as the corresponding coefficient of the $\beta$--function 
$b_i=(n_V-n_H)/6$. In the next subdivision the exponent $s_i$ 
as well the topological three point couplings 
$K_{ijk}$, the evaluation of the  second Chern class on the 
$i$'th (1,1)-form $c^2_{i}=\int_M c_2\wedge J_i$ and the
numbers of rational curves with small bi-degree are given. The
last column indicates the vanishing of elliptic curves, which was
used to fix the $r_i$.} 
\vskip 0.7cm          
Note that all the examples in table 1 have 
the same Hodge-numbers and the same
modular properties, governed by $\Gamma_0(2)_+$ \klmI, at the weak 
coupling limit in a potential dual heterotic description 
$z_s\rightarrow 0$. \foot{The present weak coupling
calculations in the $N=2$ heterotic string 
compactifications in fact
do not distinguish between them. This ambiguity will be discussed 
in detail in  \bkkII.} Their strong coupling behaviour is 
however very different as  they exhibit transitions to different 
one K\"ahler moduli cases.       
 
In the first model of table 1 as well as in the cases 4-9 
in the tables 2-4 below, the discriminant of the Picard-Fuchs 
equation exhibits, beside $z_i=z_s=0$
two (three) components 
corresponding to the conifold(s) $\Delta_c=0$ and the strong coupling 
singularity $\Delta_s=(1-4 z_s)=0$. 
Their exponents are fixed from the vanishing of the elliptic curves. 
However in these cases we have an universal contribution to the asymptotic 
behavior of $F_1^{top}$ at $\Delta_s$ from the Jacobian, which is 
$\sim \Delta_s^{1/2}$ (see below). Hence the coefficient $b_s$ is 
given by $b_s=(r_s+{1\over 2})\cdot 2$, where the factor 2 arises from 
the normalization of the $\beta$--function for $SU(2)$. 

The second model has only one discriminant component. 
$$\Delta_c=(1- z_s)^5+(1- 256 z_1)^2 + z_s z_1
(27 z_s^3-144 z_s^2+320 z_s -2816)-1.$$
However one observes that $\Delta_s=(z_s-1)=(q_s-1)+(q_s-1) f(q_s,q_1)$,
vanishes again at $t_s=0$, i.e. as before a $\IP^1$ shrinks to zero
size. Furthermore since $w_0(z)\propto \Delta_s^{-1}$ the model has 
a singular gauge choice. 
$\Delta_c\propto \Delta_s^5$ and $z_1\propto \Delta_s^5$
reflects the fact that $z_1=0$ and $\Delta_c=0$ have a point
of tangency at $t_s=0$, which has to be resolved. Adding finally 
the contribution from the Jacobian ${\rm det}\left(\partial z
\over \partial t\right)\propto \Delta_s^5$ leads to 
a\foot{Technically the situation is very similar as  
in the birational equivalent representation of the Calabi-Yau 
$N_0$ 8 by the non Fermat hypersurface 
$\cicy-240(\IP_{1,1,1,1,3}|7){2,122}$ discussed in \bkk. The second
representation has also no explicit discriminant factor $\Delta_s$
in the unresolved moduli space, but a singular gauge $w_0$ at $t_s=0$
and a proportionality of $z_1$, $\Delta_c$  and the Jacobian to $\Delta_s$ 
which add up to $\beta_s=-{2\cdot 14\over 6}$. As similar consideration
leads to $b_s=-{2\cdot 32\over 6}$ in case $N_0$ 3.}
$b_s=(3+2-(-{168\over 12}))+
5-{1\over 6}\cdot 5-{31\over 6}\cdot 5=
-{16\over 6}$, which gives precisely the difference $n_V-n_H$, which  
correspond to the inverse of the conifold transition described in \gms.
This is consistent with the interpretation of $n^{(0)}_{0,1}=16$ 
massless hyper multiplets coming from two-branes wrapping around 
the corresponding isolated curves. From the expansion of $\Delta_s$
one sees that there is no Weyl reflection in the monodromy
around $\Delta_s$. The transition is in this respect different 
from the others in table 1-3.

$$
\vbox{\offinterlineskip\tabskip=0pt 
\halign{\strut
\vrule#
&\hfil ~$#$~
&\vrule#
&\hfil ~\hfil$#$\hfil~
&\vrule#
&\hfil $\, #$~
&\hfil $#$~
&\hfil $#$~
&\vrule# 
&\hfil $\, #$~
&\hfil $#$~
&\hfil $#$~
&\hfil $#$~
&\hfil $#$~
&\hfil $#$~
&\vrule#
&\hfil ~$#$~
&\vrule#\cr
\noalign{\hrule}
& N_0&&{\rm Extremal\,\, Transition}  &&j:& r_{j}&b_{j}&&i:&s_{i}& 
K_{11i} &c^2_{i}&n^{(0)}_{i,0} &n^{(0)}_{0,i} && n^{(1)}_{i,j} &\cr
\noalign{\hrule}
&4 &&\cicy-252(\IP^4_{1,1,2,2,6}|12){2,128}       &&c:&
-{1\over 6}&-{1\over 6}        &&1:& 
-{32\over 6}&4   &52  &    2496& 2 &&n^{(1)}_{i,j}=0,&\cr
&&& \downarrow                                &&s:&
-{4\over 6} &-{2\cdot 1\over 6} &&2:&
         -3&2   &24  &  223752&   &&i<3          &\cr
&&&  \cicy-256(\IP^5_{1^5,3}|6,2){1,129}    && &
                        &                   &&1:& 
&4   & 52 &       &   &&             & \cr
\noalign{\hrule}
&5 &&\cicy-132(\IP^5_{1^2,2^4}|6,4){2,68}       &&c:&
-{1\over 6}&-{1\over 6}        &&1:& 
-{36\over 6}&12   &60  &    360& 6 &&n^{(1)}_{i,j}=0,&\cr
&&& \downarrow                                &&s:&
-{6\over 6} &-{2\cdot 3\over 6} &&2:&
         -3&6   &24  &  2682&   &&i<2          &\cr
&&&  \cicy-144(\IP^6|3,2,2){1,73}  && &
                        &                   &&1:& 
&12   & 60 &       &   &&             & \cr
\noalign{\hrule}
&6 &&  \cicy-112(\IP^6_{1^2,2^5}|4,4,4){2,58}     &&c:&
-{1\over 6}&-{1\over 6}        &&1:& 
-{19\over 6}&16   &64  &    256& 8 &&n^{(1)}_{i,j}=0,&\cr
&&& \downarrow                                &&s:&
-{7\over 6} &-{2\cdot 4\over 6} &&2:&
         -3&6   &24  &  1248&   &&i<4          &\cr
&&&  \cicy-128(\IP^7|2,2,2,2){1,65}  && &
                        &                   &&1:& 
&16   & 64 &       &   &&             & \cr
\noalign{\hrule}}
\vrule}
$$
\noindent{\bf Table 2:}  {\sl Extremal transitions starting 
from $K_3$-fibrations
at strong coupling. The modular groups at $z_s\rightarrow 0$ are 
$SL(2,\ZZ)$, $\Gamma_0(3)_+$ and $\Gamma_0(4)_+$ respectively \klmI.}

$$
\vbox{\offinterlineskip\tabskip=0pt 
\halign{\strut
\vrule#
&\hfil ~$#$~
&\vrule#
&\hfil ~\hfil$#$\hfil~
&\vrule#
&\hfil $\, #$~
&\hfil $#$~
&\hfil $#$~
&\vrule# 
&\hfil $\, #$~
&\hfil $#$~
&\hfil $#$~
&\hfil $#$~
&\hfil $#$~
&\hfil $#$~
&\vrule#
&\hfil ~$#$~
&\vrule#\cr
\noalign{\hrule}
& N_0&&{\rm Extremal\,\, Transition}  &&j:& r_{j}&b_{j}&&i:&s_{i}& 
K_{11i}&c^2_{i}&n^{(0)}_{i,0} &n^{(0)}_{0,i} && n^{(1)}_{i,j} &\cr
\noalign{\hrule}
&7 &&\cicy-144(\IP^4_{1,2,2,3,4}|12){2,74}       &&c:&
-{1\over 6}&-{1\over 6}        &&1:& 
-{22\over 6}&2   &32  &    252& 6 &&n^{(1)}_{i,j}=0,&\cr
&&& \downarrow                                &&s:&
-{6\over 6} &-{2\cdot 3\over 6} &&2:&
         -{9\over 2}&3   &42  &  -9252&   &&i<2          &\cr
&&&  \cicy-156(\IP^5_{1^3,2^2,3}|6,4){1,79}  && &
                        &                   &&1:& 
&2   & 32 &       &   &&             & \cr
\noalign{\hrule}
&8 &&\cicy-240(\IP^4_{1,2,2,2,7}|14){2,122}       &&c:&
-{1\over 6}&-{1\over 6}        &&1:& 
-{28\over 6} &2  &44  &    3& 28 &&n^{(1)}_{i,j}=0,&\cr
&&& \downarrow                                &&s:&
-{17\over 6} &-{2\cdot 14\over 6} &&2:&
            -{23\over 2}&7 &126  &   -6&   &&i<2          &\cr
&&&  \cicy-296(\IP^4_{1,1,1,1,4}|8){1,149}    && &
                        &                   &&1:& 
&2   & 44 &      &   &&             & \cr
\noalign{\hrule}}
\vrule}
$$
\noindent{\bf Table 3:} {\it  Extremal transitions from  
Calabi-Yau manifolds with a $Z_2$ singular curve, which are not 
$K_3$ fibrations\foot{For case 7.) $K_{122}=K_{222}=3$, for case 
8.) $K_{122}=21$, $K_{222}=63$.}. The inverse conifold transition
to the extremal transition $N_0$  8 was studied in \bkk .}

Let us come now to the physical interpretation of the spectrum derived
from the gravitational index. Apart from the factor of 
two, the index agrees with the difference in the Hodge numbers of 
the pairs of manifolds and fits well the interpretation of additional
massless hyper multiplets coupled to the $U(1)$ vector multiplets.
Apart from the cases where $b=0$, there is a sufficient number of
hyper multiplets to allow flat directions of the $N=2$ scalar potential.
Giving a vev to scalars in the hyper multiplets one moves
onto a new Higgs branch where the $U(1)$ is broken and the
moduli space is that of the Calabi--Yau with the corresponding, 
lower number
of vector multiplets. However we will argue that this is only half of
the story: in fact the gravitational index is 
sensitive only to the net number of vector and hyper multiplets due to
their equal contributions with opposite sign \vafaFoC. Thus the above
results fit equally well the situation where on top of the massless
hyper multiplets one has additional vector multiplets which enlarge
the $U(1)$ factors to a non-abelian gauge group $G$, accompanied 
by a hyper multiplet in the adjoint representation of $G$. Our arguments
rely on the discrete symmetries in the local monodromies, which fit
the Weyl group of $G$, and the fact that the transition are related to 
curve singularities on the Calabi--Yau manifold. A connection between
curve singularities and enhanced gauge symmetries has been discussed in 
\vafaFoC~in the local context. A similar effect in an exceptional
$N=2$ compactification with a $N=4$ characteristic spectrum 
has been investigated in \aspI.

Let us point out some features of the proposed physics of the 
singularities. 
The spectrum of a vector and hyper multiplet in the adjoint representation
plus additional matter has positive beta-function coefficient 
and a non-abelian gauge symmetry is 
not expected to be broken as in the asymptotic free case \swI.
Due to the presence of the adjoint hyper multiplet there are two 
possible Higgs breaking mechanisms to the Cartan subalgebra (CSA).
The breaking through scalars of the CSA in the vector multiplets 
corresponds to the motion in the complex structure moduli space
of the mirror $\Xh$. However to obtain gauge symmetry enhancement
we have also to tune to zero the values of the scalars in the 
hyper multiplets, which belong to the K\"ahler moduli and can not be 
represented by polynomial deformations of $\ph$. 
Therefore the factorization of K\"ahler and complex structure
moduli spaces fails due to the gauge couplings between the
vector and hyper multiplets.

\newsec{Curve singularities}
It turns out that one type of singularities we want to understand
is related to curve singularities in the Calabi--Yau manifold. In
particular such singularities can arise if a Calabi--Yau
is described as the subvariety of a singular ambient space, as happens
to be the case in the example \xtwelve. The weighted projective space
has quotient singularities, if subsets of the weights have a non-trivial
factor $n$ in common; according to whether the Calabi--Yau variety intersects
the singular set of the ambient space in a point or a curve, its 
singularities  are locally of the type $\IC^3/\ZZ_n$ or $\IC^2/\ZZ_n$,
respectively.

The resolution of the ambient space quotient singularity 
is standard and can be described in terms of toric geometry;
we refer to the literature for details \resol. In toric language
the singular curve $C$ corresponds to a one-dimensional edge of the dual 
polyhedron with integral lattice points on it. 
The resolution process adds a new vertex for each
of these points and to each of these vertices corresponds an exceptional
$\IP^1$ bundle over $C$ in the blown up of the Calabi--Yau manifold.
For a $\ZZ_n$ curve singularity the intersection matrix of the exceptional
$\IP^1$' s is proportional to the Cartan matrix of $A_{n-1}$ 
with self-intersections normalized to $-1$.\br
\goodbreak\midinsert
\centerline{\epsfxsize 2.3truein\epsfbox{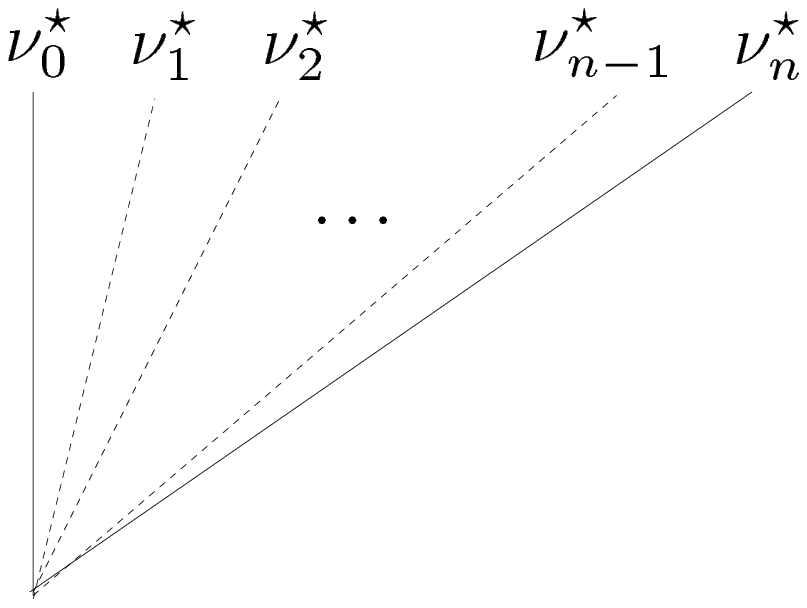}}
\smallskip
\leftskip 2pc \rightskip 2pc
\noindent{
\ninepoint\sl \baselineskip=8pt
{\bf Fig.2}
The toric resolution of the $\ZZ_n$ singularity. To each vertex
$\nu^\star_i$ corresponds a perturbation $X_1^{n-i}X_2^i$ in the 
defining polynomial.}
\smallskip\endinsert
In virtue of the monomial-divisor map the new vertices correspond to 
the addition of new perturbations to the original defining polynomial
of $\Xh$. The fact that the integral points are arranged on a one-dimensional
line translates to the following form of the perturbations:

\def\nus{\nu^\star}
\eqn\perts{
X_1^{n},X_1^{n-1}X_2^{1},\ \dots,X_2^{n}\ ,
}%
where $X_1$ and $X_2$ denote monomials in those projective coordinates, 
which are zero on the singular curve and the first and last perturbations
correspond to the original vertices $\nus_0$ and $\nus_n$, respectively.
The set of perturbations \perts\ will lead to singular configurations
of the mirror Calabi--Yau $\Xh$ for special values of the moduli 
$\phi_0...\phi_N$ of the new perturbations. In particular there is a
point in the moduli space where 
\eqn\csI{
\sum_k \phi_k x^{n-k}y^{k} = (x+y)^n \ .
}%
It is important to note that the singularities of the defining polynomial
$\ph$ appear in the complex structure moduli space, whereas the volume
of the exceptional divisors of the blow up is controlled by K\"ahler moduli,
which are not described by polynomial deformations of $\Xh$. Thus the 
relation to the original quotient singularity, 
leading to the singular curve
on $X$, is not immediately obvious. However it happens that the 
singular locus in the complex structure moduli space of $\Xh$ maps
by the mirror map to regions in the K\"ahler moduli space of $X$, corresponding to 
one or more of the exceptional divisors being blown down to a point.

Let us investigate in more detail the simplest quotient singularity 
$\ZZ_2$. This is the case appearing in the Calabi--Yau manifold \xtwelve, 
which we will use as a representative example, although the arguments
apply more generally.
The locus of the quotient singularity is
$x_1=x_2=0$, which is a fixed point of the 
projective action $x_i \to \lambda^{w_i}x_i$ with $\lambda=-1$. On  
$\Xh$ there is a singular curve described by
\eqn\singcur{
x_3^6+x_4^6+x_5^2 = 0\ .
}%
The resolution process adds a new vertex $\nus_1$ to the dual simplex,
which is the average of two of the generic vertices $\nus_0,\nus_2$ 
of the singular ambient space, 
\eqn\newvec{\nus_1=\h(\nus_0+\nus_2)}
In the polynomial $\ph$ of \xtwelve,
the vertices $\nus_0,\ \nus_2$ are related by the 
monomial-divisor mirror map to the polynomials $x_1^{12},\ x_2^{12}$,
while the blow up perturbation related to $\nus_1$ is represented
by $x_1^6x_2^6$. Obviously $\Xh$ becomes singular for the 
special value $\phi=\pm 1$ where $\ph$ contains the perfect square
$(x_1^6\pm x_2^6)$; this singularity is the origin of the 
discriminant factor $\Delta_s$. 

The $\ZZ_2$ singularity of the ambient space results in the 
singular locus $\phi^2=1$ in the complex structure moduli space of the
mirror manifold $\Xh$. More interestingly, the mirror map related to 
the complex structure modulus of the blow up, $z_s$, has the property
observed in \klmI:
\eqn\mm{
0=\Delta_s=z_s(t_1,t_s)-{1\over 4} \ \Leftrightarrow \ t_s=0,
}%
independently of the value of $t_1$. Therefore $\Delta_s=0$ maps
on the type IIA side to a face of the K\"ahler cone with the
corresponding $b$ field set to zero. 
The resolution of the $\ZZ_2$ singularity in the context of transitions
through phase boundaries of the complexified K\"ahler moduli space has
been discussed in \agm. The special coordinate $t_s$ is used to define
the area of Riemannian surfaces which lie in the exceptional divisor $E_s$.
That is the image of $\Delta_s=0$ on the type IIB side 
corresponds to the blow down of $E_s$ on $X$ and the Calabi--Yau 
develops the original curve singularity. A relation between curve 
singularities and non-perturbative enhanced gauge 
symmetries has been established in \refs{\bsv} in a local 
analysis.

The behavior \mm\ of the mirror map can be traced back to 
two basic properties of the toric data of the Calabi--Yau
variety. The first are those generators of the
Mori cone, which express the relations between the vertices arising
from the blow up. For $n=2$ we have a single linear relation, \newvec,
which is represented by a vector $l_s$,
\eqn\morivec{
l_s=(1,1,-2,0,\ \dots,0) \ .
}%
In \agm\ it was shown, how $l$ determines the ordinary hypergeometric
differential operator ${\cal L}(f)=\theta_z^2f-z\theta_z(\theta_z+1/2)f,\ 
(\theta\equiv z d/dz)$, 
which governs the mirror map between
$t_s$ and $z_s$ on a special rational curve, parametrized by $z_s$, 
in the compactified moduli space. In a Calabi--Yau phase the 
rational curve is defined by setting the remaining complex structure
moduli to the large complex structure point, $z_i=0$. The relevant
solution of the differential equation is \agm
\eqn\perI{
t_s={1\over 2\pi i}
\ln\Big({1-2z_s-2\sqrt{1-4z_s}\over 2z_s} \Big) \ .
}%
First note that $t_s$ vanishes for $z_s={1\over 4}$. 
Secondly, under the motion
of $z$ once around $z={1\over 4}$, $t_s$ transforms as $t_s \to -t_s$.
The period $t_s$ is a special coordinate of the large complex structure
limit; in fact it will correspond to the heterotic dilaton in 
the cases where the dual description exist. A particle charged only
w.r.t. the corresponding $U(1)$ factor becomes massless at $t_s=0$;
moreover it is interpreted as an {\it electrically} charged excitation 
w.r.t. 
the basis chosen in the large complex point (possibly
associated to a weakly coupled heterotic theory). 

To extend our arguments from $z_i=0$ to the whole codimension one
locus $\Delta_s=0$, assume that the discriminant factor of the 
CY contains the factor $\Delta_s=z_s-{1\over 4}$. It is clear the the 
local monodromy around $\Delta_s=0$ and thus also the leading behavior of
the periods in an expansion around $\epsilon=z-{1\over 4}$ cannot depend on the 
remaining moduli $z_i$. Therefore $t_s \sim \epsilon^{1/2}$ all
along $\Delta_s=0$ and \mm\ holds\foot{Actually it is important for this
kind of argument that the topology of the singular locus $\Delta_s$ 
does not change when the remaining moduli are switched on. This happens 
in the case of the conifold discriminant $\Delta_{c}$ in \disI.}.
 
We can give now also an explanation for the overall shift of $\h$,
which we had to take into account to extract the spectrum from the
coefficient $r_s$ of $F_1^{top}$, as the contribution 
of the Jacobian of the mirror map. Namely, at $t_s=0$, not only
$z_s=0$, but also its first derivative:
\eqn\jacI{
{\d z_s \over \d t_s}\sim t \sim (z_s-{1\over 4})^{\h}\sim \Delta_s^{\h} \ .}

The differential operator ${\cal L}$ is simply the restriction of
one operator of the complete Picard--Fuchs system to the locus $z_i=0$. 
In the $n=2$ case it is not difficult to understand \mm\ as a consequence
of the fact, that the $z_s$ dependence of the period $t_s$ is in general 
described by a series of hypergeometric functions, not only for $z_i=0$.
This follows from the structure of the generators of the Mori cone 
which are essentially of the form
\def\lsb{\bar{l}_s}
\eqn\morigen{
\eqalign{
l_s&=\ (1,1,-2,0,\dots,0)\ ,\cr
\lsb&=\ (0,0,1,\dots)\ ,\cr
l_k&=\ (0,0,0,\dots) \ .}}
where the $l_s,\lsb,l_k$ correspond to the complex structure moduli $z_s,\bar{z}_s,z_k$.
There can be  modifications of the precise expression for $\lsb$, the
important point being that there is a single vector $\lsb$ which overlaps
with the vector $\l_s$ of \morivec. If there would be more
relations involving the vertices $\nus_i,\  i=0..2$, of the $\ZZ_2$ 
singularity, it would mean that this singularity is part of a more 
complicated singularity. In this case we can still switch off the
perturbations corresponding to the vertices $\nus_j$ involved in the
additional relations to get a situation as in \morigen; in fact
we will use such a procedure below to generalize from the $\ZZ_2$
to the $\ZZ_n$ case.

From \morigen\ and the expression for the fundamental 
period in the large complex structure limit \Yau\ ,
it follows that the $z_s$ dependence
can be described in terms of hypergeometric functions:
\eqn\fper{
\omega_0=\sum_{\bar{n},n_1,\dots,n_k} F(\bar{n},n_1,...,n_k)
\bar{z}^{\bar{n}}z_1^{n_1}...z_k^{n_k}\ u_{\bar{n}}(z_s),
}%
where $u_\mu(z_s)=(z_s)^{-\mu/2}  
{_2 F _1}(-\h \mu, -\h \mu + \h, 1; 1/z_s)$ for the precise choice of $\lsb$ in 
\morigen. This kind of representation of the fundamental period has been 
introduced in \Candelas\  , to determine the full period vector and 
the mirror maps for the two $K_3$ fibrations 
$\cicy-168(\IP^4_{1,1,2,2,2}|8){2,86}$
and
$\cicy-252(\IP^4_{1,1,2,2,6}|12){2,128}$. 
For the discussion of the strong coupling singularity
we focus on the $z_s$ dependence which is more general; in the following
we will use heavily the results and notations of ref. \Candelas.
In fact the only information we need is
that the $z_s$ dependence of {\it all} the periods related to the special 
coordinates, is described by two series of function, namely $u_n$ and
the second linear independent solution to the hypergeometric equation,
called $v_n$ in \Candelas. From the transformation behavior of $t_s$ under the local 
monodromy it follows that the $z_s$ dependence of $t_s$ is proportional to
$v_n=u_nv_0-(4z_s)^{-\h}\sqrt{1-4z_s}f_n({1\over z_s})$, 
where $f_n$ is a polynomial
and $v_0=2\ln\big((4z_s)^{-\h}(1+\sqrt{1-4z_s})\big)-i\pi$.
The term proportional to $v_0$ gives rise to the limiting behavior \perI,
while the second term controls the part which depends on the remaining moduli:
\eqn\thfin{
t_s = -{1\over \pi i}\ln\big((4z_s)^{-\h}(1+\sqrt{1-4z_s})\big)+
z_s^{-\h}\sqrt{1-4z_s}F(z_s,z_k) \ .
}%
Inversion yields $z_s = (q_s-1)^2 f(q_s,q_1)$.

To generalize to $\ZZ_n$ we use first the fact that each triple of successive vertices
$\nus_{i-1},\ \nus_i,\ \nus_{i+1}$ fulfils the same  relation  as the vertices 
in \newvec\
and can be used to reach a limiting point described by the mirror map \perI. This 
corresponds
to the possibility to blow down each single $\IP^1_i$ of the resolution while
keeping the remaining volumes arbitrarily large. 
\goodbreak\midinsert
\centerline{\epsfxsize 2.3truein\epsfbox{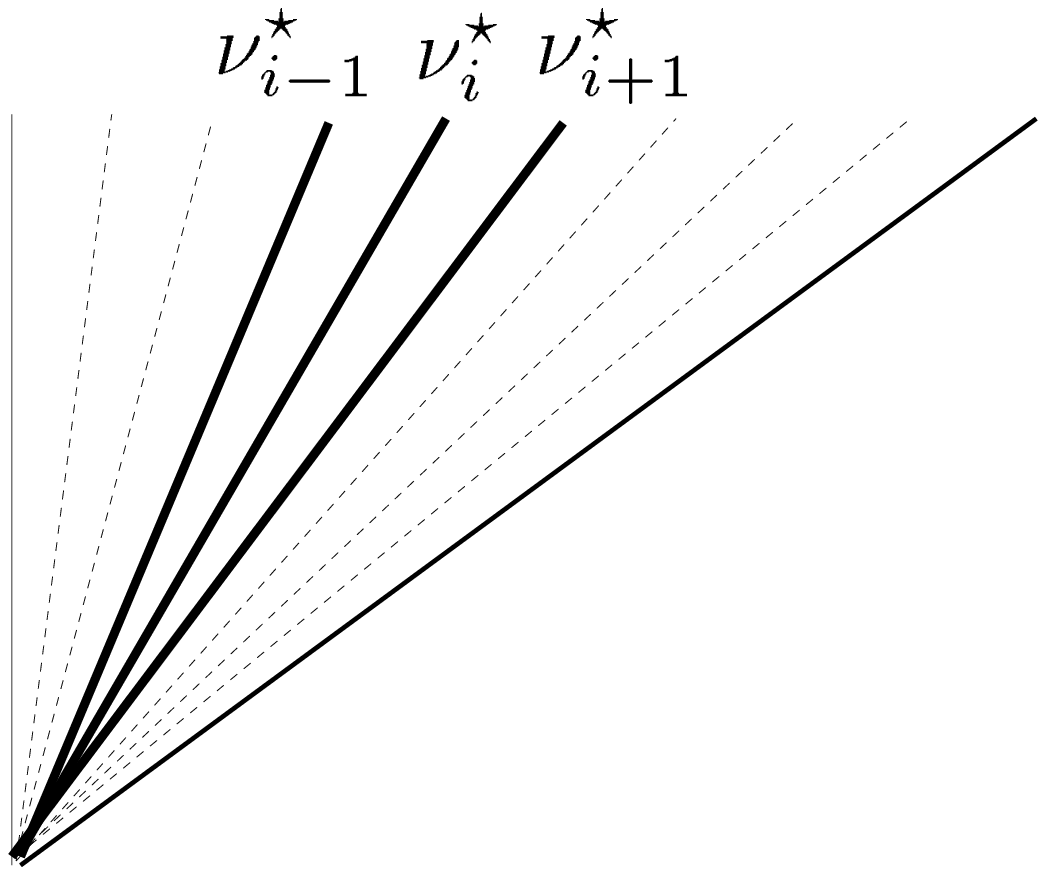}}
\smallskip
\leftskip 2pc \rightskip 2pc
\noindent{
\ninepoint\sl \baselineskip=8pt
{\bf Fig.2}
Three vertices of the blow up of the $\ZZ_n$ singularity defining 
a limit where one exceptional $\IP^1_i$ is blown down to a point.}
\smallskip\endinsert
In this way we obtain $n-1$ 
reflections
$\sigma_i:\ t_s^i \to -t_s^i$ acting on the special coordinates $t^i_s$. The global
properties are governed by the intersection form $\Omega$
of the exceptional divisors $E_i$ of 
complex codimension one, which are $\IP^1$ bundles over the singular curve $C$ fixed
by the $\ZZ_n$ action. The quadratic intersection form of the $E_i$ defined by
$J \cdot E_i \cdot E_j$, where $J$ is the K\"ahler form of the ambient space,  
descends from the intersection form of the rational curves and is given by 
\dais\hktyII:
\eqn\iform{
J\cdot E_j^2=-2\Lambda,\qquad J\cdot E_i\cdot E_j = \Lambda,\  |i-j|=1,\qquad
J\cdot E_i\cdot E_j = 0,\  {\rm otherwise} \ ,
}%
where $\Lambda$ is a numerical factor which does not matter in the present context.
The full monodromy group of the $\ZZ_n$ discriminant factor is generated by the
reflections $\sigma_i$ and has to leave invariant the intersection form $\Omega$
which is proportional to the 
Cartan matrix of $A_{n-1}$ in virtue of \iform.
What we have described above are precisely the properties of the Weyl group
of $SU(n)$ ! In fact the Weyl transformations can be recovered from 
the reflections and the products at the intersections by induction in $n$.

A nice illustration of how the gravitational index manages to fit the 
enhancement at the intersection of two $SU(2)$ factors is given by the
second and third three moduli examples of table 4. The $\ZZ_3$ discriminant
factor generalizing $\Delta_s=1-4z_s$ for $\ZZ_2$ is 
\eqn\disII{
\Delta_s=1-4z_s^1-4z_s^2+18z_s^1z_s^2-27(z_s^1z_s^2)^2\ .}
The differential equation governing the behavior of the mirror map relating
the special coordinates $t_s^i$ to the algebraic moduli $z_s^i$ is
determined by two vectors of the Mori cone
\eqn\morII{
l_s^1=(0,1,1,-2,0,\dots,0),\ l_s^2=(1,0,-2,1,0,\dots,0)}
and the solution of the corresponding differential equation at the limiting 
$SU(2)$ points $z_s^i=0,\ z_s^j={1\over 4}$ 
and the intersection point $z_s^1=z_s^2={1\over 3}$
are given by \perI\ and 
\eqn\solx{
t_1 = t_2 = \ln\Big({1-z_s-\sqrt{1-2z_s-3z_s^2}\over 2z_s}\Big) \ ,\ \ 
z_s\equiv z_s^1=z_s^2 \ ,} respectively.

On the $SU(2)$ lines the situation parallels the case of the $\ZZ_2$ singularities
described above.  At the intersection point $z_s^1=z_s^2=1/3$ we expect 
an additional number of massless hyper multiplets fitting in representations
of $SU(3)$. At this points the discriminant factor \disII\ is singular
and becomes proportional to $\epsilon^2$ instead of $\epsilon$, where
$\epsilon \sim \ z_s-1$. Similarly the Jacobian contains a factor of
$\epsilon$ instead of $\epsilon^{\h}$, due to the fact that two of the three rows 
are $\sim \epsilon^{\h}$. Finally in order to get the correct normalization
for $SU(3)$ we have to rescale by a factor of 3 instead of 2. The resulting formula
$b_s=(2r_s+1)\times 3$ matches precisely the number of rational curves
lying in the exceptional divisors of the blow down, shown in table 4. 
The number of hyper multiplets fits precisely into adjoint representations of $SU(3)$.

In view of the universal contribution of $\h$ from the Jacobian for the
case of $SU(2)$ and the jump at the intersection point of two $SU(2)$ factors
it is suggestive to think about the opposite contributions of the Jacobian
and the discriminant factors to the gravitational index as the opposite 
contribution of massless vector and hyper multiplets. Such an interpretation is
supported by the by the cancellation appearing in the first example of table 4,
where we find $b_s=(-1/2+1/2)\times 2=0$ for the discriminant factor $\Delta_s=
(1-4z_s)$. 
The spectrum fitting these data is just an adjoint vector and hyper multiplet
without additional matter. This spectrum can be further confirmed by calculating
the monodromy and the periods in the local coordinates vanishing at $\Delta_s$,
where one finds 
that the two characteristic periods with eigenvalue minus one show no
logarithmic behavior.

$$
\vbox{\offinterlineskip\tabskip=0pt 
\halign{\strut
\vrule#
&\hfil ~$#$~
&\vrule#
&\hfil ~\hfil$#$\hfil~
&\vrule#
&\hfil $\, #$~
&\hfil $#$~
&\hfil $#$~
&\vrule# 
&\hfil $\, #$~
&\hfil $#$~
&\hfil $#$~
&\hfil $#$~
&\hfil $#$~
&\hfil $#$~
&\vrule#
&\hfil ~$#$~
&\vrule#\cr
\noalign{\hrule}
& N_0&&{\rm CY\,\, manifold}  &&j:& r_{j}&b_{j}&&i:&s_{i}& 
c^2_{i}&n^{(0)}_{i,0,0} &n^{(0)}_{0,i,0}& n^{(0)}_{0,0,i} 
&& n^{(1)}_{i,j,k} &\cr
\noalign{\hrule}
&9 &&\cicy-480(\IP^4_{1,1,2,8,12}|24){3,243}       &&c_1:&
-{1\over 6}&-{1\over 6}        &&1:& 
-{52\over 6}        &92  &    480& 0 &-2 &&n^{(1)}_{i,j,k}=0,&\cr
&&&                               &&s:&
-{3\over 6} &-{2\cdot 0 \over 6} &&2:&
-3&24  & 480 &0   & 0 &&i<2          &\cr
&  &&                                              &&c_2: &
-{1\over 6} &-{1\over 6}&&3:& 
-5& 48& 480 & 0 &0 &&             & \cr
\noalign{\hrule}
&10 &&\cicy-132(\IP^4_{1,2,3,3,3}|12){3,69}       &&c:&
-{1\over 6}&-{1\over 6}        &&1:& 
-5        &48  &    56& 4 &4 &&n^{(1)}_{i,j,k}=0,&\cr
&&&                               &&s:&
-{5\over 6} &-{2\cdot 2\over 6} &&2:&
-{17\over 3}&56  & -272 &0   & 0 &&i<2          &\cr
&  &&                                              && &
&&&3:&-3 & 24& 3240 & 0 &0 &&             & \cr
\noalign{\hrule}
&11 &&\cicy-192(\IP^4_{1,2,3,3,9}|18){3,99}       &&c:&
-{1\over 6}&-{1\over 6}        &&1:& 
-{27\over 6}        &42  &    252& 2 &2 &&n^{(1)}_{i,j,k}=0,&\cr
&&&                               &&s:&
-{4\over 6} &-{2\cdot 1\over 6} &&2:&
-3&24  & -9252 &0   & 0 &&i<2          &\cr
&  &&                                              && &
&  &&3:& 
-{16\over 3}& 52&84862& 0 &0 &&             & \cr
\noalign{\hrule}
}
\vrule}
$$
\noindent{\bf Table 4:} {\it Three moduli $K_3$--fibration 
Calabi-Yau manifolds. The first has a $Z_2$ singular curve
with an exceptional $Z_4$ point. The last two have $Z_3$ 
singular curves.}

\newsec{Comments}
We have analyzed two types of transitions between topological different
Calabi--Yau manifolds, 
which have a physical interpretation in terms of new branches
in the moduli space arising from additional massless degrees of freedom. 
The first describes the inverse transition of the conifold transition 
described in \gms. The second one is related to curve singularities
on the Calabi--Yau manifold and corresponds to  a situation with enhanced non-abelian
gauge symmetries with a non asymptotically free spectrum. They have in common the
fact that on the transition point the volume of certain rational curves becomes 
zero. In the examples where a dual heterotic description exists, the
singularities lie in the strongly coupled regime. This provides a new,
stringy realization of non-abelian gauge symmetries through solitonic 
degrees of freedom in a $N=2$ theory at strong string coupling,
invisible in the perturbation theory.\br\br
{\bf Acknowledgements}\br
We would like to thank Per Berglund, Sheldon Katz, Wolfgang Lerche 
and Fernando Quevedo for helpful discussions.

\listrefs
\end